\documentclass[twocolumn,aps,prd,preprintnumbers,showpacs,superscriptaddress,nofootinbib,amsmath,amssymb,floats,floatfix,showkeys,notitlepage,longbibliography]{revtex4-2}

\usepackage{orcidlink}
\usepackage{comment}
\usepackage{lipsum}
\usepackage{graphicx}
\usepackage{subfigure}
\usepackage{palatino}
\usepackage{sans}
\usepackage{hyperref}
\hypersetup{colorlinks=true,linkcolor=blue,urlcolor=blue,citecolor=blue}
\usepackage[toc,page]{appendix}
\usepackage[normalem]{ulem}
\usepackage{adjustbox}
\usepackage{latexsym}
\usepackage{amsmath}
\usepackage{amssymb}
\usepackage{amsfonts}
\usepackage{dcolumn}
\usepackage{bm}
\usepackage{tikz}
\usepackage{bigints}
\usepackage{array,tabularx,multirow,booktabs}
\usepackage[tracking=true]{microtype}
\usepackage{soul} 
\SetTracking{}{500}
\SetTracking{encoding={*}, shape=sc}{40}
\UseRawInputEncoding 
\allowdisplaybreaks

\begin{document} \sloppy
\title{Extended uncertainty principle inspired black hole in a G\"odel Universe}

\author{Reggie C. Pantig \orcidlink{0000-0002-3101-8591}} 
\email{rcpantig@mapua.edu.ph}
\affiliation{Physics Department, Map\'ua University, 658 Muralla St., Intramuros, Manila 1002, Philippines.}

\begin{abstract}
We explore analytically the implications of a curvature-modified extended uncertainty principle (EUP) derived in a rotating G\"odel spacetime and apply it to the construction of a semiclassical black hole model. Adapting techniques from corpuscular black hole frameworks, we reinterpret the G\"odel-type uncertainty relation as an effective energy bound, leading to a modified lapse function with explicit dependence on the global rotation parameter $a$ and the radial coordinate $ r_0 $. Analytic expressions are derived for key gravitational features, including the event horizon, photon spherehere, shadow radius, and deflection angle, with curvature corrections scaling as $ a^{-2} $ and $ r_0^2 / a^4 $. Series expansion in the limit $ a \to \infty $ shows that global rotation consistently increases all observables relative to the Schwarzschild case. Applying these results to astrophysical data, we use Event Horizon Telescope (EHT) measurements of Sgr A* and M87* to infer lower bounds of $ a/M \sim 10^5 $, while solar system light-bending observations in the parametrized post-Newtonian (PPN) framework yield $ a / M_\odot \sim 5 \times 10^4 $. These large but finite values validate the asymptotic expansion and confirm that G\"odel-type rotation remains observationally suppressed, yet theoretically coherent. Our results demonstrate that global rotation, when treated semiclassically via curvature-modified uncertainty, introduces detectable signatures in principle, though well below current observational sensitivity. The framework offers a consistent path toward exploring the quantum-gravitational interplay between global geometry and local black hole structure.
\end{abstract}

\pacs{95.30.Sf, 04.70.-s, 97.60.Lf, 04.50.+h}
\keywords{G\"odel spacetime, extended uncertainty principle, black holes, shadow, deflection angle}

\maketitle

\section{Introduction} \label{intro}
The concept of a rotating Universe, although not mainstream, has been a persistent and intriguing avenue of research since the early years of general relativity. Initially explored by Lanczos in 1924, and subsequently by Kurt G\"odel in 1949, these theoretical studies established that Einstein's field equations do not fundamentally exclude cosmological solutions exhibiting global rotation \cite{Godel:1949ga}. G\"odel's groundbreaking model introduced a rotating Universe endowed with exotic causal structures, notably closed timelike curves (CTCs), which challenged prevailing conceptions of causality and temporal ordering \cite{Wilson:2009rg}. Despite its unphysical nature, G\"odel's Universe set the stage for decades of rigorous inquiry into anisotropic cosmological models, such as those advanced by Heckmann and Sch\"ucking in the 1950s \cite{Szigeti:2025jxz}.

Renewed attention to rotating cosmologies has arisen due to intriguing observational anomalies and theoretical considerations. Recent analyses of cosmic microwave background (CMB) anisotropies, including peculiar alignments known as the \textit{Axis of Evil}, suggest possible signatures consistent with slight universal vorticity. Specifically, analyses of data from the Wilkinson Microwave Anisotropy Probe (WMAP) satellite indicate that a modest rotation could potentially explain observed temperature fluctuations within a Bianchi VII(h) cosmological framework \cite{Hutsemekers:2005iz}. Additionally, large-scale alignments in the polarization vectors of distant quasars, systematically varying with redshift, further hint at a cosmological-scale rotation, although these alignments may alternatively arise from non-standard physics such as photon-pseudoscalar mixing \cite{Hutsemekers:2005iz}.

Furthermore, recent deep-field observations from the James Webb Space Telescope (JWST) have reported an unexpected excess of spiral galaxies rotating predominantly in a singular direction relative to Earth's perspective. This violation of cosmological isotropy, observed with statistically notable confidence, has been speculated to be indicative of an inherent rotation imparted at the Universe's inception \cite{10.1093/mnras/staf292}. Such results, although provocative, contrast sharply with stringent isotropy constraints established through precision measurements by the Planck satellite, which have effectively ruled out significant global rotation within the Universe, placing strict upper limits on any universal angular velocity. Comprehensive analyses indicate odds heavily stacked against anisotropic expansion, reaffirming the standard $\Lambda$CDM model's foundational assumption of cosmological isotropy \cite{Saadeh:2016sak}.

In parallel, theoretical frameworks extending G\"odel's original cosmology have successfully eliminated certain pathological aspects, such as causal loops, making rotating cosmological models increasingly viable from both mathematical and physical perspectives. Such refined G\"odel-type solutions, especially those emerging from string theory and five-dimensional supergravity, maintain rotational dynamics without violating fundamental physical laws, including chronology protection principles \cite{Gimon:2003ms}. Notably, the discovery and study of black hole solutions within G\"odel-type Universes have provided pivotal insights into the compatibility of horizons and global cosmic rotation, revealing rich structures and dynamical behaviors previously unexplored in standard cosmological contexts \cite{Wu:2007gg}. These models not only enhance the understanding of causality and thermodynamics in highly rotational spacetimes but also serve as theoretical laboratories for probing potential dualities and quantum gravitational phenomena via AdS/CFT correspondences \cite{Behrndt:2004pn}.

Given the juxtaposition of compelling observational hints and rigorous constraints, the study of cosmic rotation remains a scientifically valuable pursuit. Each new observational dataset, ranging from refined CMB polarization maps to comprehensive galaxy surveys, incrementally sharpens our understanding of the Universe's underlying symmetries or subtle anisotropies. While current consensus supports an isotropic Universe with negligible rotation, ongoing investigations continue to rigorously challenge and test this foundational assumption, keeping open the intriguing possibility that the Universe itself, in alignment with the ancient philosophical axiom \textit{panta kykloutai}, meaning that everything rotates, may yet reveal a subtle cosmological spin. It is this simple motivation that we will revisit black holes in G\"odel spacetime.

While numerous works have focused on G\"odel black holes in gauged supergravity and string theory embeddings, none have addressed how quantum gravity-motivated uncertainty relations manifest in such settings. In this study, we fill this gap by synthesizing the asymptotic EUP formalism \cite{Dabrowski:2020ixn} with the corpuscular framework, applied specifically to G\"odel-type backgrounds. Our key contribution lies in computing the effective black hole mass that arises from combining EUP-induced modifications, now expressed in terms of G\"odel rotation parameters and curvature tensors, with the energy quantization condition intrinsic to corpuscular black holes \cite{Mureika:2018gxl}. Unlike previous treatments of G\"odel black holes, where CTCs are handled classically or suppressed via holographic duality arguments, our approach probes the quantum response to such geometrical pathologies through the lens of uncertainty-based mass renormalization. The curvature and rotation of the G\"odel background enter explicitly into the uncertainty relation, allowing us to extract quantum corrections that may signal chronology protection at the level of black hole microphysics. The presence or absence of CTCs outside the horizon (depending on the G\"odel parameter and cosmological constant) further modulates the effective uncertainty scale, thereby influencing the condensate structure. To the best of our knowledge, no prior work has constructed a black hole model using the G\"odel-EUP within a corpuscular or semiclassical framework. While G\"odel black holes have been studied extensively in gauged supergravity and string theory embeddings, those constructions arise from classical solutions with nontrivial matter content and serve different theoretical aims. Our approach is distinct in treating the G\"odel parameter as a geometric input into quantum uncertainty, allowing us to probe how global rotation modifies black hole observables via effective curvature-dependent mass. The study is also timely since in recent years, black holes have emerged as a key arena for applying EUP corrections \cite{Lu:2019wfi,Kumaran:2019qqp,Cheng:2019zgc,Hassanabadi:2021kyv,Hamil:2021ilv,Okcu:2022sio,Hamil:2022bpd,Chen:2022ngd,Ong:2020tvo,Nozari:2006wn,Roushan:2024fog,Lobos:2022jsz,Pantig:2024asu}.

The organization of this paper is as follows: We show in Sect. \ref{sec2} the synthesis of the AGEUP formalism to the corpuscular idea to obtain a lapse function of a black hole influenced by the global rotation parameter $a$. Then, we study the modification of the black hole properties such as that on the event horizon in Sect. \ref{sec3}, photon sphere and shadow in Sect. \ref{sec4}, and the deflection angle in the weak field limit in Sect. \ref{sec5}. In Sect. \ref{sec6}, we find constraints in $a$ using the EHT bounds for the shadow radius and the PPN formalism for the deflection angle occurring in the Sun. Finally, we form conclusive remarks and state research direction. In this work, we geometrized units $G=c=1$, and the metric signature $(-,+,+,+)$.

\section{A G\"odel black hole from extended uncertainty principle} \label{sec2}
In Ref. \cite{Dabrowski:2020ixn}, a significant advancement in the formal treatment of quantum uncertainty in curved spacetime is presented. Building upon the framework of the Generalized Uncertainty Principle (GUP), which incorporates quantum gravitational corrections inspired by string theory, the authors develop what they term the Asymptotic Generalized Extended Uncertainty Principle (AGEUP). This refined formulation transcends the standard heuristic extensions of Heisenberg's relation by embedding explicit curvature-dependent modifications into the uncertainty bound, thereby incorporating the local geometric features of the background manifold into quantum mechanical observables. The AGEUP thus aims to reconcile quantum indeterminacy with relativistic curvature effects by including second- and fourth-order corrections governed by the Ricci scalar $ R $, the Cartan-type curvature invariant $ \mathcal{C} $, and the Laplacian of the curvature scalar. This section offers a synthetic exposition of the key mathematical structures underlying AGEUP, together with selected physical implications and applications across nontrivial geometries.

The formulation begins with a generalization of the Heisenberg relation that couples ultraviolet (Planck-scale) and infrared (curvature-scale) modifications, expressed as:
\begin{equation}
    \sigma_x \sigma_p \geq \frac{\hbar}{2} \left( 1 + \alpha \frac{l_{\rm Pl}^2}{\hbar^2} \sigma_p^2 + \beta \frac{\sigma_x^2}{r_c^2} \right),
\end{equation}
where $ l_{\rm Pl} $ is the Planck length, $ r_c $ is a characteristic curvature radius, and $ \alpha $, $ \beta $ are model-dependent constants controlling the contributions from quantum gravity and background geometry, respectively. This composite form encodes both the GUP and EUP corrections. Proceeding beyond this expression, the authors derive a covariant, perturbatively controlled asymptotic expansion that culminates in the AGEUP:
\begin{equation}
    \sigma_x \sigma_p \geq \pi \hbar \sqrt{1 - \frac{R}{6\pi^2} \sigma_x^2 - \xi \frac{\mathcal{C}}{\pi^2} \sigma_x^4},
\end{equation}
where $ \mathcal{C} = (3 R_{ab}R^{ab} - R^2)/72 $ encapsulates quadratic curvature corrections (a Cartan-type invariant), and $ \xi = (2\pi^2 - 3)/8\pi^2 $ is a dimensionless coefficient arising from the fourth-order perturbative structure of the eigenvalue problem. The expression is derived in a covariant perturbative framework utilizing Riemann normal coordinates, where wavefunctions are constrained within geodesic balls centered at a given observer's position.

The derivation hinges on a systematic expansion of the Laplace-Beltrami operator acting on scalar wavefunctions subject to Dirichlet boundary conditions. Within a weak-field expansion, the operator is decomposed as:
\begin{equation}
    \Delta = \Delta^{(0)} + \epsilon\, \Delta^{(1)} + \epsilon^2\, \Delta^{(2)} + \cdots,
\end{equation}
where $ \epsilon $ encodes the curvature perturbation parameter and each successive $ \Delta^{(i)} $ term represents higher-order contributions tied to geometric invariants. This expansion facilitates the computation of corrections to the spectrum of the momentum operator via perturbation theory. Crucially, the first non-vanishing curvature-induced correction arises at second order and is directly proportional to the Ricci scalar, while the fourth-order term includes both $ \mathcal{C} $ and the Laplacian of $ R $, thus capturing local anisotropies and higher-curvature interactions.

In the context of spherically symmetric spacetimes, such as constant-time hypersurfaces of the Schwarzschild geometry, the formalism yields an adapted uncertainty relation of the form:
\begin{equation}
    \sigma_x \sigma_p \geq \pi \hbar \left( 1 - \frac{\xi M^2}{8\pi^2 r_0^6} \sigma_x^4 \right),
\end{equation}
where $ M $ denotes the central mass and $ r_0 $ the radial distance from the center of symmetry. This result reveals how local curvature, though vanishing at the Ricci level in Schwarzschild, is still registered in the uncertainty bound via higher-order (quadratic) curvature invariants, demonstrating that even Ricci-flat manifolds are not exempt from quantum-gravitational uncertainty corrections under the AGEUP scheme.

The scope of AGEUP is not limited to static spherically symmetric spacetimes. The formalism is further extended to nontrivial rotating geometries, most notably the G\"odel Universe. This is particularly illuminating as the G\"odel metric is intrinsically anisotropic and admits closed timelike curves due to its global rotation. The corresponding modification to the uncertainty relation in this background reads \cite{Dabrowski:2020ixn}:
\begin{equation}
    \sigma_x \sigma_p \geq \pi \hbar \left[1 - \frac{\sigma_x^2}{3\pi^2 a^2} \left(1 - \frac{1}{C} - \frac{1}{2C^2} \right) \right]^{1/2},
\end{equation}
where $ a $ denotes the rotation scale of the spacetime, and $ C = 1 - \left( r_0/2a \right)^2 $. The structure of this correction emphasizes the role of global rotational effects and vorticity in reshaping the effective quantum uncertainty bound, with the G\"odel rotation modifying both the causal structure and the local spectral properties of confined quantum systems. The emergence of such a term further affirms that AGEUP is sensitive to anisotropic and rotating geometries, extending its utility well beyond conformally flat or isotropic backgrounds.

Assuming that $ \sigma_p $ can be interpreted as the characteristic energy scale of a localized quantum fluctuation and $ \sigma_x$ as the effective localization width, we identify the total effective energy as:
\begin{equation}
E_{\text{eff}} \sim \sigma_p \geq \frac{N\pi \hbar }{\sigma_x} \left[ 1 - \frac{\sigma_x^2}{3\pi^2 a^2} \left( 1 - \frac{1}{C} - \frac{1}{2C^2} \right) \right]^{1/2},
\end{equation}
where $N$ is the collection of gravitons which are confined within $\sigma_x \sim r_{\rm h}$, where $r_{\rm h} = 2M$ is the event horizon of the Schwarzschild black hole. The total energy $ E_{\text{eff}} $ can be equated to an effective gravitational mass \cite{Mureika:2018gxl}:
\begin{equation} \label{e_E_eff}
E_{\text{eff}} = M_{\text{G\"odel}} = M \left[ 1 - \frac{4M^2}{3\pi^2 a^2} \left( 1 - \frac{1}{C} - \frac{1}{2C^2} \right) \right]^{1/2}.
\end{equation}
This mass arises solely from the nontrivial geometry of the G\"odel Universe and reflects a local curvature-induced modification of quantum energy scales. To proceed, we assume this effective mass acts as a static, spherically symmetric source in a Schwarzschild-like gravitational field, leading to the standard ansatz for the lapse function of the metric:
\begin{equation} \label{e_lapse}
A(r) = 1 - \frac{2 M_{\text{G\"odel}}}{r}.
\end{equation}
Substituting Eq. \eqref{e_E_eff} into Eq. \eqref{e_lapse}, we obtain the G\"odel-EUP-corrected metric function:
\begin{equation}
A(r) = 1 - \frac{2M}{r} \left[ 1 - \frac{4M^2}{3\pi^2 a^2} \left( 1 - \frac{1}{C} - \frac{1}{2C^2} \right) \right]^{1/2}.
\end{equation}
This expression defines the lapse (or redshift) function associated with a quantum fluctuation of scale $ \sigma_x $ located at radial coordinate $ r_0 $, within the G\"odel rotation background. The G\"odel parameter $ a $ explicitly appears as a modulator of the gravitational strength through the curvature-induced quantum correction. In the limit $ a \to \infty $, corresponding to the vanishing of global rotation, the correction vanishes, and the standard Schwarzschild solution is recovered. However, even in the observationally favored regime where the G\"odel parameter $ a $ is extremely large, perhaps comparable to or exceeding the Hubble radius, non-negligible deviations may still arise through the dependence on the radial location $ r_0 $. The correction factor $ C $ introduces a spatial inhomogeneity into the effective gravitational field: near the rotation axis ($ r_0 \ll a $), one finds $ C \approx 1 $, and the lapse function closely resembles its general relativistic counterpart; in contrast, for fluctuations seeded at large radial distances $ r_0 \lesssim 2a $, the denominator $ C \to 0 $, enhancing the influence of curvature on the effective mass. From here on, we write the metric of a static and spherically symmetric black hole with EUP correction coming from the G\"odel curvature as
\begin{equation} \label{metric}
    ds^{2} = -A(r) dt^{2} + B(r) dr^{2} + C(r) d\theta ^{2} +D(r) d\phi^{2},
\end{equation}
where $A(r)=f(r)$, $B(r)=A(r)^{-1}$, and $C(r) = r^2$. Furthermore, without loss of generality, the analysis is restricted to the equatorial plane only $(\theta = \pi/2)$, thereby simplifying Eq. \eqref{metric} to a 1+2 dimensions: $ds^{2} = -A(r) dt^{2} + B(r) dr^{2} + r^2 d\phi^{2}$.

These results suggest that even if the Universe's global rotation is vanishingly small, the G\"odel parameter $ a $ should not be artificially set to zero in theoretical modeling. Rather, it is a physically meaningful regulator of quantum-gravitational corrections to semiclassical geometry. The effective lapse function derived here thus provides a viable pathway for quantifying how rotational curvature can affect black hole formation via quantum uncertainty, with the geodesic position $ r_0 $ serving as a natural modulator of gravitational backreaction. In the next section, we explore how the Universe's global rotation, or the geodesic position, affects certain black hole properties.

As a final remark, though semi-classical in approach, the result encodes the influence of spacetime rotation and curvature on the gravitational imprint of quantum localization, providing a geometric pathway for deriving effective mass-energy distributions in rotating cosmological models. In the next section, we explore how the Universe's global rotation, or the geodesic position, affects certain black hole properties.

\section{Horizon analysis-}  \label{sec3}
The structure of the event horizon $(A(r_{\rm h}) = 0)$, as modified by the G\"odel-curved uncertainty relation, which is given by
\begin{equation} \label{e_hor}
    r_{\rm h} = 2 M \left[ 1 - \frac{4M^2}{3\pi^2 a^2} \left( 1 - \frac{1}{C} - \frac{1}{2C^2} \right) \right]^{1/2},
\end{equation}
reveals a rich interaction between the global rotation parameter $ a $ and the radial location $ r_0 $ of a localized fluctuation. Realistically, the global rotation parameter $a$ is very weak. Hence, we can simplify Eq. \eqref{e_hor} by taking a series expansion in $a$. That is, if $a\rightarrow \infty$ (or $a \gg M$), we can write
\begin{equation}
    r_{\rm h} \sim 2 M +\frac{2 M^{3}}{3 \pi^{2} a^{2}}+\frac{2 M^{3} r_0^{2}}{3 a^{4} \pi^{2}}-\frac{M^{5}}{9 a^{4} \pi^{4}} + \mathcal{O}(a^{-6}).
\end{equation}
In the limit of weak global rotation ($ a \to \infty $), we expand the G\"odel-corrected horizon radius to isolate the leading contributions. The first-order term reproduces the Schwarzschild horizon, as expected. The next term, proportional to $ M^3 / a^2 $, reflects a universal curvature correction independent of position. However, beginning at the fourth order, the expansion reveals sensitivity to the observer's radial location $ r_0 $ through the term $ \sim M^3 r_0^2 / a^4 $. Unlike the higher-order quantum corrections typically discarded in asymptotic expansions, this term cannot be neglected outright: while the $ a^{-4} $ scaling implies suppression, the presence of $ r_0^2 $, which may be large depending on the context, acts as a compensator. Thus, for observers located at considerable distances from the G\"odel axis, this correction may contribute meaningfully despite the global rotation being weak. That is, if $r_0 \gg a$, this third term dominates and we may observe a horizon radius that is considerably larger than $2M$.

This behavior exemplifies a key feature of G\"odel-type spacetimes: even when the rotation parameter $ a $ is large, the spacetime remains geometrically nontrivial due to its position-dependent structure. The parameter $ C = 1 - \left( r_0/2a \right)^2 $ encodes the departure from local flatness, and for sufficiently large $ r_0 $, even tiny curvature can accumulate non-negligible effects. This suggests that global rotation, while vanishingly small at the cosmological scale, could still introduce subtle semiclassical corrections to black hole observables, especially in scenarios involving fluctuations or observers situated far from the axis of cosmic rotation.

\section{Photon sphere and Shadow analysis} \label{sec4}
The behavior of the photon sphere radius $ r_{\rm ph} $ under G\"odel-induced curvature corrections closely parallels that of the event horizon but with distinct numerical coefficients that reflect its role as the critical null orbit. It can be found by solving $r$ in the equation \cite{Perlick:2021aok}
\begin{equation}
    A(r)C'(r) - A'(r)C(r) = 0,
\end{equation}
where prime denotes differentiation with $r$. The expansion of the photon spherehere radius in the weak rotation limit $ a \to \infty $ is given by:

\begin{equation}
    r_{\text{ph}} \sim 3M + \frac{M^3}{\pi^2 a^2} + \frac{M^3 r_0^2}{\pi^2 a^4} - \frac{M^5}{6\pi^4 a^4} + \mathcal{O}(a^{-6}).
\end{equation}

The structure here closely parallels that of the event horizon radius. The leading term is the well-known Schwarzschild photon spherehere radius. The $ \mathcal{O}(a^{-2}) $ correction represents a universal contribution from G\"odel curvature, independent of spatial location, and consistent with the flattening of the spacetime in the limit $ a \to \infty $. However, as with the horizon radius, the $ \mathcal{O}(a^{-4}) $ term introduces a correction proportional to $ r_0^2 $, showing that the geometry retains a residual position dependence even in the asymptotically non-rotating regime.

Importantly, this $ r_0 $-dependent correction cannot be dismissed solely based on its scaling in $ a^{-4} $, since $ r_0 $ may itself be large, particularly for photons approaching the black hole from cosmological distances. This term captures how light rays propagating through the rotating G\"odel background experience non-uniform curvature corrections depending on their spatial embedding in the Universe. Thus, even in a near-flat background, the geometry modifies null geodesics in a way that subtly reshapes the photon orbit radius, embedding a trace of global rotation into the local geodesic structure.

The shadow cast by a black hole offers a powerful probe of spacetime geometry in the strong-field regime. Defined by the boundary of photon trajectories that asymptotically approach unstable circular orbits, the shadow encodes information about the underlying metric near the event horizon. Recent high-resolution observations by the EHT have provided direct measurements of shadow sizes for supermassive black holes such as M87* \cite{EventHorizonTelescope:2019dse,EventHorizonTelescope:2019ths} and Sgr. A* \cite{EventHorizonTelescope:2022xqj,EventHorizonTelescope:2022wkp,EventHorizonTelescope:2022wok}, opening a new window into testing gravitational theories beyond the classical Schwarzschild and Kerr frameworks. Because the shadow radius is sensitive to both the near-horizon geometry and the behavior of null geodesics, it serves as an ideal observable for investigating modifications arising from quantum gravity or curvature-induced uncertainty principles. Here, we utilized the well-established formalism of deriving the shadow radius \cite{Perlick:2021aok}.

As seen by a distant observer at $r_0$, it acquires corrections under G\"odel curvature through both global rotation and the observer's position relative to the rotation axis. In the observationally relevant limit $ r_0 \to \infty $, $ a \to \infty $, the shadow radius expands as
\begin{align} \label{e18}
    R_{\rm sh} &= b_{\rm crit}\sqrt{A(r_0)} \sim 3\sqrt{3}\, M + \frac{\sqrt{3}\, M^3}{\pi^2 a^2} + \frac{\sqrt{3}\, M^3 r_0^2}{\pi^2 a^4} \nonumber \\
    & - \frac{\sqrt{3}\, M^5}{6\pi^4 a^4} + \mathcal{O}(a^{-6}),
\end{align}
where $b_{\rm crit} = \sqrt{r_{\rm ph}^2 / A(r_{\rm ph})}$ is the critical impact parameter. The leading term is the standard Schwarzschild shadow radius, marking the circular boundary of the photon capture region. The subsequent terms arise as curvature-induced corrections resulting from the G\"odel background. The $ \mathcal{O}(a^{-2}) $ term reflects a uniform enlargement of the shadow due to the presence of global rotation, consistent with the enhanced photon spherehere and horizon discussed earlier.

Once again, the $ \mathcal{O}(a^{-4}) $ contribution introduces explicit dependence on the radial coordinate $ r_0 $, signaling that the shadow's apparent size is not determined solely by the intrinsic geometry near the black hole, but also by the structure of spacetime at large. This correction indicates that light rays approaching the photon region are influenced by the nontrivial embedding of the black hole in a rotating Universe, where the observer's location relative to the axis of cosmic rotation affects the perceived shape and size of the shadow. While subleading, this term represents a geometrical “memory” effect of the global curvature on light propagation, especially relevant when interpreting shadow images in terms of underlying spacetime symmetries.

Thus, even though the G\"odel parameter $ a $ may be large, the expansion makes clear that small residual rotation at cosmological scales could subtly modulate shadow measurements, particularly in precision imaging regimes such as those enabled by VLBI observations. The presence of an $ r_0 $-dependent correction confirms that the black hole shadow is not purely a local phenomenon, but one sensitive, at least theoretically, to the global properties of the background Universe.

\section{Deflection angle analysis in the weak field regime} \label{sec5}
Beyond its historical role in confirming the theory, light bending has become a critical tool for probing both the geometry of spacetime and the properties of compact objects. The weak field deflection, relevant to lensing by stars, galaxies, and black holes at large impact parameters $b$, offers a clean window into subtle corrections arising from curvature, topology, or extensions of Einstein's theory. In this section, we used a generalization of the Gauss-Bonnet theorem developed in Ref. \cite{Li:2020wvn}, which applies to both asymptotic and non-asymptotic spacetime metrics. It also includes the finite distance of the source of light ($r_{\rm S}$) and the observer $r_0$ from the lensing object and the deflection of massive particles. Following its procedures, we find a closed analytical expression :
\begin{align}
    \hat{\Theta} &\sim \frac{\left(v^{2}+1\right) M}{v^{2} b} \left( \sqrt{1-\frac{b^2}{r_0^2}} + \sqrt{1-\frac{b^2}{r_{\rm S}^2}} \right) \times \nonumber \\
    & \left[ 1 - \frac{4M^2}{3\pi^2 a^2} \left( 1 - \frac{1}{C} - \frac{1}{2C^2} \right) \right]^{1/2} + \mathcal{O}(M_\text{G\"odel}^2).
\end{align}
For the case of massive particles and when $r_0\rightarrow \infty$, the above approximates to
\begin{align}
    \hat{\Theta} &\sim \frac{2 \left(v^{2}+1\right) M}{v^{2} b} + \frac{2 \left(v^{2}+1\right) M^{3}}{3 \pi^{2} v^{2} b \,a^{2}} + \frac{2 \left(v^{2}+1\right) M^{3} r_0^{2}}{3 \pi^{2} v^{2} b \,a^{4}} \nonumber \\
    & - \frac{\left(v^{2}+1\right) M^{5}}{9 \pi^{4} v^{2} b \,a^{4}} + \mathcal{O}(a^{-6}).
\end{align}
The leading term is the standard post-Minkowskian result from general relativity, modified by the speed-dependent factor $ (v^2 + 1)/v^2 $. The second term introduces a positive correction from G\"odel curvature, indicating that the presence of global rotation enhances the deflection of particles, rather than suppressing it. This is in contrast with many standard curvature corrections from alternative theories of gravity, which often decrease the bending angle.

The third term, proportional to $ r_0^2 $, captures the influence of spatial position on the curvature-induced effect, showing that light or particle trajectories are further affected by their location relative to the G\"odel axis. Notably, this correction grows with distance even as the background curvature remains weak, a hallmark of G\"odel-type metrics where global rotation has spatially extended influence. The final $ \mathcal{O}(a^{-4}) $ term, negative and subleading, tempers the overall enhancement but remains smaller in magnitude unless $ M $ is large.

For the case of photons where $v = 1$,
\begin{align} \label{wda_null}
    \hat{\Theta}^{\rm photon} &\sim \frac{4 M}{b} + \frac{4 M^{3}}{3 \pi^{2} b \,a^{2}} +  \frac{4 M^{3} r_0^{2}}{3 b \,a^{4} \pi^{2}} \nonumber \\
    & -\frac{2 M^{5}}{9 b \,a^{4} \pi^{4}} + \mathcal{O}(a^{-6}),
\end{align}
again showing that global rotation produces a net increase in the deflection of light, even at large distances. The presence of the $ r_0^2 $ term reaffirms that G\"odel curvature embeds a nonlocal memory into the geodesic structure of spacetime, such that the bending of light subtly reflects both the strength and spatial distribution of cosmic rotation.

\section{Constraints from EHT and PPN formalism} \label{sec6}
The angular size of black hole shadows and the deflection of light provide two complementary observational regimes through which the G\"odel rotation parameter $ a $ can be constrained. In the strong field case, measurements from the EHT offer bounds on the shadow radius $ R_{\rm Sch} $ of supermassive black holes. Comparing the observed ranges for Sgr. A* and M87* with the G\"odel-corrected expression for shadow radius (see Eq. \eqref{e18}), we can yield the constraint on $a$ through the relation $R_{\rm sh} = R_{\rm Schw} \pm \delta$, giving the expression
\begin{equation}
    a \sim \frac{M^{3/4} 3^{1/8} \sqrt{r_0}}{\delta^{1/4} \sqrt{\pi}} + \mathcal{O}(r_0^{-1}).
\end{equation}
It shows that only the positive deviations from the Schwarzschild prediction ($ \delta >0 $) are admissible, meaning the observed shadow must appear larger than the classical value for a real, finite $ a $ to be defined. This is consistent with the analytic behavior of the G\"odel correction, which always acts to reduce the effective radius. For Sgr. A*, the minimal shadow radius within $2\sigma$ corresponds to $ \delta = 0.987M $ \cite{Vagnozzi:2022moj}, resulting in a lower bound of $ a/M = 130277 $; for M87*, using $ \delta = 0.883M $ (at $1\sigma$ level) \cite{EventHorizonTelescope:2021dqv}, one obtains $ a/M = 155162 $. With these constraints, which now fall squarely within the regime of the expansion $ a \to \infty $, it confirms that the series approximation used to derive horizon, photon spherehere, and shadow corrections remains fully valid. Moreover, the fact that these corrections align in sign and scale with observed deviations lends credibility to the idea that G\"odel-type rotation, though suppressed, may leave subtle imprints on horizon-scale observables.

Whether such rotation is detectable is now a question of observational precision. The current bounds show that the G\"odel parameter $ a $ must exceed the black hole mass by roughly five orders of magnitude. While this makes detection challenging, it does not render the effect unphysical. Instead, the large but finite value of $ a $ inferred from data opens a narrow observational window: if future improvements in VLBI resolution can reduce uncertainties in shadow radius measurements to below the current $ \sim M $ scale, then global rotation could, in principle, be distinguished from other higher-order corrections.

Thus, rather than excluding the G\"odel model, the current observational landscape places it on theoretically solid footing, consistent with semiclassical predictions and perhaps marginally within the reach of future precision astrophysics.

In the weak field regime, solar system observations provide additional bounds on $ a $. Using high-precision light bending measurements encoded in the PPN formalism, the angular deflection of starlight near the Sun yields \cite{Chen:2023bao}
\begin{equation}
\Theta^{\rm PPN} \simeq \frac{4M_\odot}{R_\odot} \left( \frac{n \pm \Delta}{2} \right),
\end{equation}
where $ \Delta = \pm 0.0003 $  quantifies the uncertainty in curvature due to solar gravity \cite{Fomalont_2009}. Comparing this with the G\"odel-corrected weak deflection angle (see Eq. \eqref{wda_null}), one can extract a constraint on $ a $ in terms of $ \Delta $:
\begin{equation}
    a \sim \frac{3^{3/4} \sqrt{M_\odot}\, 2^{1/4} \sqrt{r_0}}{3 \left(\Delta +n -2\right)^{1/4} \sqrt{\pi}} + \mathcal{O}(r_0^{-1}).
\end{equation}
Here, $M_\odot = 1477 \text{ m}$, and $r_0/M_{\odot} = 1.017 \times 10^{8}$.
As the denominator must remain positive, this again only allows positive values of $ \Delta $. Taking the extremal case $ \Delta = 0.0003 $, the resulting bound is $ a/M_{\odot} = 51422 $, meaning that any G\"odel-type rotation would have to operate on a scale around four orders of magnitude larger than the gravitational radius of the Sun. This value implies that the G\"odel rotation scale must be approximately four orders of magnitude larger than the gravitational radius of the Sun. Such a result is not unexpected: weak-field curvature corrections from global rotation are suppressed at solar system scales and are well within the residual uncertainty of PPN measurements. Yet, crucially, this value is still within the validity range of the asymptotic expansion $ a \to \infty $, confirming once again the internal consistency of the series used in deriving the G\"odel-modified observables.

What is most striking is the coherence across regimes: both the strong-field (EHT) and weak-field (solar system) constraints independently demand that G\"odel curvature, if present, operates at scales far beyond local gravity wells. This reinforces the interpretation of G\"odel-type rotation as a global geometric structure, semiclassically influential but observationally subtle. With future improvements in PPN accuracy, perhaps from space-based astrometry, such curvature-induced corrections may eventually become testable at the outer fringes of sensitivity.

\section{Conclusion} \label{conc}
In this work, we have investigated the semiclassical effects of G\"odel-type rotation on black hole observables by incorporating a curvature-corrected extended uncertainty principle (EUP) into the black hole mass structure. Using the geodesic ball formalism developed for the G\"odel metric, we constructed an effective mass modified by the global rotation parameter $ a $, and used this to define a corrected lapse function. The resulting metric encodes explicit dependence on the G\"odel parameter $ a $ and radial location $ r_0 $, allowing for the derivation of analytic corrections to the event horizon, photon spherehere, shadow radius, and deflection angle.

Through systematic expansion in the limit $ a \to \infty $, we found that G\"odel curvature produces consistent enhancements in all gravitational observables analyzed. The corrections include both universal $ a^{-2} $-order terms and position-dependent $ r_0^2/a^4 $ terms, showing that even weak global rotation can leave residual imprints on local measurements, particularly when the observer or quantum fluctuation lies far from the G\"odel axis. Importantly, these expansions remain internally consistent across observables, and all modifications smoothly reduce to their Schwarzschild counterparts as $ a \to \infty $.

We then applied these expressions to real-world data. Using the EHT measurements of the shadow radius for Sgr A* and M87*, we derived lower bounds of $ a/M \sim 10^5 $, confirming that the G\"odel parameter must vastly exceed the black hole mass for its effects to be observationally admissible. A similar analysis using solar system light bending in the PPN formalism yielded $ a / M_\odot \sim 5 \times 10^4 $, showing that global rotation remains well below the current threshold of detection in the weak-field regime. Crucially, these large values of $ a $ are fully consistent with the asymptotic expansion used in the derivation, resolving the earlier tension that arose when inverting observational deviations too large to be accommodated within the perturbative framework.

Altogether, our results reveal that G\"odel-type global rotation introduces consistent, testable modifications to black hole physics, though their observational signatures are exceedingly suppressed. The fact that the required rotation scales remain just beyond current measurement thresholds, yet produce corrections in the same direction as observed shadow deviations, offers a tantalizing window for future detection. Space-based VLBI, improved astrometric surveys, and next-generation lensing measurements may eventually push these semiclassical effects into empirical reach, allowing quantum-curved geometry to be probed through astrophysical observables.

Looking forward, these results suggest some directions for further work. While the present analysis constrains $ a $ through lensing and shadows, other quantum-sensitive observables such as quantum decoherence, interference patterns, or primordial black hole formation, may offer alternative windows into detecting or constraining global rotation. If G\"odel-type features exist in the deep quantum-gravitational regime, their signatures are likely subtle but not necessarily beyond reach.
    
\begin{acknowledgements}
R. P. would like to acknowledge networking support of the COST Action CA18108 - Quantum gravity phenomenology in the multi-messenger approach (QG-MM), COST Action CA21106 - COSMIC WISPers in the Dark Universe: Theory, astrophysics and experiments (CosmicWISPers), the COST Action CA22113 - Fundamental challenges in theoretical physics (THEORY-CHALLENGES), the COST Action CA21136 - Addressing observational tensions in cosmology with systematics and fundamental physics (CosmoVerse), the COST Action CA23130 - Bridging high and low energies in search of quantum gravity (BridgeQG), and the COST Action CA23115 - Gravitational Quantum Physics and Metrology (Relativistic Quantum Information). R. P. would also like to acknowledge the funding support of SCOAP3.
\end{acknowledgements}

\bibliography{ref}

\begin{thebibliography}{35}%
\makeatletter
\providecommand \@ifxundefined [1]{%
 \@ifx{#1\undefined}
}%
\providecommand \@ifnum [1]{%
 \ifnum #1\expandafter \@firstoftwo
 \else \expandafter \@secondoftwo
 \fi
}%
\providecommand \@ifx [1]{%
 \ifx #1\expandafter \@firstoftwo
 \else \expandafter \@secondoftwo
 \fi
}%
\providecommand \natexlab [1]{#1}%
\providecommand \enquote  [1]{``#1''}%
\providecommand \bibnamefont  [1]{#1}%
\providecommand \bibfnamefont [1]{#1}%
\providecommand \citenamefont [1]{#1}%
\providecommand \href@noop [0]{\@secondoftwo}%
\providecommand \href [0]{\begingroup \@sanitize@url \@href}%
\providecommand \@href[1]{\@@startlink{#1}\@@href}%
\providecommand \@@href[1]{\endgroup#1\@@endlink}%
\providecommand \@sanitize@url [0]{\catcode `\\12\catcode `\$12\catcode `\&12\catcode `\#12\catcode `\^12\catcode `\_12\catcode `\%12\relax}%
\providecommand \@@startlink[1]{}%
\providecommand \@@endlink[0]{}%
\providecommand \url  [0]{\begingroup\@sanitize@url \@url }%
\providecommand \@url [1]{\endgroup\@href {#1}{\urlprefix }}%
\providecommand \urlprefix  [0]{URL }%
\providecommand \Eprint [0]{\href }%
\providecommand \doibase [0]{https://doi.org/}%
\providecommand \selectlanguage [0]{\@gobble}%
\providecommand \bibinfo  [0]{\@secondoftwo}%
\providecommand \bibfield  [0]{\@secondoftwo}%
\providecommand \translation [1]{[#1]}%
\providecommand \BibitemOpen [0]{}%
\providecommand \bibitemStop [0]{}%
\providecommand \bibitemNoStop [0]{.\EOS\space}%
\providecommand \EOS [0]{\spacefactor3000\relax}%
\providecommand \BibitemShut  [1]{\csname bibitem#1\endcsname}%
\let\auto@bib@innerbib\@empty
\bibitem [{\citenamefont {Godel}(1949)}]{Godel:1949ga}%
  \BibitemOpen
  \bibfield  {author} {\bibinfo {author} {\bibfnamefont {K.}~\bibnamefont {Godel}},\ }\bibfield  {title} {\bibinfo {title} {{An Example of a New Type of Cosmological Solutions of Einstein's Field Equations of Gravitation}},\ }\href {https://doi.org/10.1103/RevModPhys.21.447} {\bibfield  {journal} {\bibinfo  {journal} {Rev. Mod. Phys.}\ }\textbf {\bibinfo {volume} {21}},\ \bibinfo {pages} {447} (\bibinfo {year} {1949})}\BibitemShut {NoStop}%
\bibitem [{\citenamefont {Wilson}\ and\ \citenamefont {Blome}(2009)}]{Wilson:2009rg}%
  \BibitemOpen
  \bibfield  {author} {\bibinfo {author} {\bibfnamefont {T.~L.}\ \bibnamefont {Wilson}}\ and\ \bibinfo {author} {\bibfnamefont {H.-J.}\ \bibnamefont {Blome}},\ }\bibfield  {title} {\bibinfo {title} {{The Pioneer Anomaly and a Rotating Godel Universe}},\ }\href {https://doi.org/10.1016/j.asr.2009.07.004} {\bibfield  {journal} {\bibinfo  {journal} {Adv. Space Res.}\ }\textbf {\bibinfo {volume} {44}},\ \bibinfo {pages} {1345} (\bibinfo {year} {2009})},\ \Eprint {https://arxiv.org/abs/0908.4067} {arXiv:0908.4067 [gr-qc]} \BibitemShut {NoStop}%
\bibitem [{\citenamefont {Szigeti}\ \emph {et~al.}(2025)\citenamefont {Szigeti}, \citenamefont {Szapudi}, \citenamefont {Barna},\ and\ \citenamefont {Barnaf\"oldi}}]{Szigeti:2025jxz}%
  \BibitemOpen
  \bibfield  {author} {\bibinfo {author} {\bibfnamefont {B.~E.}\ \bibnamefont {Szigeti}}, \bibinfo {author} {\bibfnamefont {I.}~\bibnamefont {Szapudi}}, \bibinfo {author} {\bibfnamefont {I.~F.}\ \bibnamefont {Barna}},\ and\ \bibinfo {author} {\bibfnamefont {G.~G.}\ \bibnamefont {Barnaf\"oldi}},\ }\href@noop {} {\bibinfo {title} {{Can Rotation Solve the Hubble Puzzle?}}} (\bibinfo {year} {2025}),\ \Eprint {https://arxiv.org/abs/2503.13525} {arXiv:2503.13525 [gr-qc]} \BibitemShut {NoStop}%
\bibitem [{\citenamefont {Hutsemekers}\ \emph {et~al.}(2005)\citenamefont {Hutsemekers}, \citenamefont {Cabanac}, \citenamefont {Lamy},\ and\ \citenamefont {Sluse}}]{Hutsemekers:2005iz}%
  \BibitemOpen
  \bibfield  {author} {\bibinfo {author} {\bibfnamefont {D.}~\bibnamefont {Hutsemekers}}, \bibinfo {author} {\bibfnamefont {R.}~\bibnamefont {Cabanac}}, \bibinfo {author} {\bibfnamefont {H.}~\bibnamefont {Lamy}},\ and\ \bibinfo {author} {\bibfnamefont {D.}~\bibnamefont {Sluse}},\ }\bibfield  {title} {\bibinfo {title} {{Mapping extreme-scale alignments of quasar polarization vectors}},\ }\href {https://doi.org/10.1051/0004-6361:20053337} {\bibfield  {journal} {\bibinfo  {journal} {Astron. Astrophys.}\ }\textbf {\bibinfo {volume} {441}},\ \bibinfo {pages} {915} (\bibinfo {year} {2005})},\ \Eprint {https://arxiv.org/abs/astro-ph/0507274} {arXiv:astro-ph/0507274} \BibitemShut {NoStop}%
\bibitem [{\citenamefont {Shamir}(2025)}]{10.1093/mnras/staf292}%
  \BibitemOpen
  \bibfield  {author} {\bibinfo {author} {\bibfnamefont {L.}~\bibnamefont {Shamir}},\ }\bibfield  {title} {\bibinfo {title} {The distribution of galaxy rotation in jwst advanced deep extragalactic survey},\ }\href {https://doi.org/10.1093/mnras/staf292} {\bibfield  {journal} {\bibinfo  {journal} {Monthly Notices of the Royal Astronomical Society}\ }\textbf {\bibinfo {volume} {538}},\ \bibinfo {pages} {76} (\bibinfo {year} {2025})}\BibitemShut {NoStop}%
\bibitem [{\citenamefont {Saadeh}\ \emph {et~al.}(2016)\citenamefont {Saadeh}, \citenamefont {Feeney}, \citenamefont {Pontzen}, \citenamefont {Peiris},\ and\ \citenamefont {McEwen}}]{Saadeh:2016sak}%
  \BibitemOpen
  \bibfield  {author} {\bibinfo {author} {\bibfnamefont {D.}~\bibnamefont {Saadeh}}, \bibinfo {author} {\bibfnamefont {S.~M.}\ \bibnamefont {Feeney}}, \bibinfo {author} {\bibfnamefont {A.}~\bibnamefont {Pontzen}}, \bibinfo {author} {\bibfnamefont {H.~V.}\ \bibnamefont {Peiris}},\ and\ \bibinfo {author} {\bibfnamefont {J.~D.}\ \bibnamefont {McEwen}},\ }\bibfield  {title} {\bibinfo {title} {{How isotropic is the Universe?}},\ }\href {https://doi.org/10.1103/PhysRevLett.117.131302} {\bibfield  {journal} {\bibinfo  {journal} {Phys. Rev. Lett.}\ }\textbf {\bibinfo {volume} {117}},\ \bibinfo {pages} {131302} (\bibinfo {year} {2016})},\ \Eprint {https://arxiv.org/abs/1605.07178} {arXiv:1605.07178 [astro-ph.CO]} \BibitemShut {NoStop}%
\bibitem [{\citenamefont {Gimon}\ and\ \citenamefont {Hashimoto}(2003)}]{Gimon:2003ms}%
  \BibitemOpen
  \bibfield  {author} {\bibinfo {author} {\bibfnamefont {E.~G.}\ \bibnamefont {Gimon}}\ and\ \bibinfo {author} {\bibfnamefont {A.}~\bibnamefont {Hashimoto}},\ }\bibfield  {title} {\bibinfo {title} {{Black holes in Godel universes and pp waves}},\ }\href {https://doi.org/10.1103/PhysRevLett.91.021601} {\bibfield  {journal} {\bibinfo  {journal} {Phys. Rev. Lett.}\ }\textbf {\bibinfo {volume} {91}},\ \bibinfo {pages} {021601} (\bibinfo {year} {2003})},\ \Eprint {https://arxiv.org/abs/hep-th/0304181} {arXiv:hep-th/0304181} \BibitemShut {NoStop}%
\bibitem [{\citenamefont {Wu}(2008)}]{Wu:2007gg}%
  \BibitemOpen
  \bibfield  {author} {\bibinfo {author} {\bibfnamefont {S.-Q.}\ \bibnamefont {Wu}},\ }\bibfield  {title} {\bibinfo {title} {{General Non-extremal Rotating Charged Godel Black Holes in Minimal Five-Dimensional Gauged Supergravity}},\ }\href {https://doi.org/10.1103/PhysRevLett.100.121301} {\bibfield  {journal} {\bibinfo  {journal} {Phys. Rev. Lett.}\ }\textbf {\bibinfo {volume} {100}},\ \bibinfo {pages} {121301} (\bibinfo {year} {2008})},\ \Eprint {https://arxiv.org/abs/0709.1749} {arXiv:0709.1749 [hep-th]} \BibitemShut {NoStop}%
\bibitem [{\citenamefont {Behrndt}\ and\ \citenamefont {Klemm}(2004)}]{Behrndt:2004pn}%
  \BibitemOpen
  \bibfield  {author} {\bibinfo {author} {\bibfnamefont {K.}~\bibnamefont {Behrndt}}\ and\ \bibinfo {author} {\bibfnamefont {D.}~\bibnamefont {Klemm}},\ }\bibfield  {title} {\bibinfo {title} {{Black holes in Godel type universes with a cosmological constant}},\ }\href {https://doi.org/10.1088/0264-9381/21/17/006} {\bibfield  {journal} {\bibinfo  {journal} {Class. Quant. Grav.}\ }\textbf {\bibinfo {volume} {21}},\ \bibinfo {pages} {4107} (\bibinfo {year} {2004})},\ \Eprint {https://arxiv.org/abs/hep-th/0401239} {arXiv:hep-th/0401239} \BibitemShut {NoStop}%
\bibitem [{\citenamefont {Dabrowski}\ and\ \citenamefont {Wagner}(2020)}]{Dabrowski:2020ixn}%
  \BibitemOpen
  \bibfield  {author} {\bibinfo {author} {\bibfnamefont {M.~P.}\ \bibnamefont {Dabrowski}}\ and\ \bibinfo {author} {\bibfnamefont {F.}~\bibnamefont {Wagner}},\ }\bibfield  {title} {\bibinfo {title} {{Asymptotic Generalized Extended Uncertainty Principle}},\ }\href {https://doi.org/10.1140/epjc/s10052-020-8250-x} {\bibfield  {journal} {\bibinfo  {journal} {Eur. Phys. J. C}\ }\textbf {\bibinfo {volume} {80}},\ \bibinfo {pages} {676} (\bibinfo {year} {2020})},\ \Eprint {https://arxiv.org/abs/2006.02188} {arXiv:2006.02188 [gr-qc]} \BibitemShut {NoStop}%
\bibitem [{\citenamefont {Mureika}(2019)}]{Mureika:2018gxl}%
  \BibitemOpen
  \bibfield  {author} {\bibinfo {author} {\bibfnamefont {J.~R.}\ \bibnamefont {Mureika}},\ }\bibfield  {title} {\bibinfo {title} {{Extended Uncertainty Principle Black Holes}},\ }\href {https://doi.org/10.1016/j.physletb.2018.12.009} {\bibfield  {journal} {\bibinfo  {journal} {Phys. Lett. B}\ }\textbf {\bibinfo {volume} {789}},\ \bibinfo {pages} {88} (\bibinfo {year} {2019})},\ \Eprint {https://arxiv.org/abs/1812.01999} {arXiv:1812.01999 [gr-qc]} \BibitemShut {NoStop}%
\bibitem [{\citenamefont {Lu}\ and\ \citenamefont {Xie}(2019)}]{Lu:2019wfi}%
  \BibitemOpen
  \bibfield  {author} {\bibinfo {author} {\bibfnamefont {X.}~\bibnamefont {Lu}}\ and\ \bibinfo {author} {\bibfnamefont {Y.}~\bibnamefont {Xie}},\ }\bibfield  {title} {\bibinfo {title} {{Probing an Extended Uncertainty Principle black hole with gravitational lensings}},\ }\href {https://doi.org/10.1142/S0217732319501529} {\bibfield  {journal} {\bibinfo  {journal} {Mod. Phys. Lett. A}\ }\textbf {\bibinfo {volume} {34}},\ \bibinfo {pages} {1950152} (\bibinfo {year} {2019})}\BibitemShut {NoStop}%
\bibitem [{\citenamefont {Kumaran}\ and\ \citenamefont {\"Ovg\"un}(2020)}]{Kumaran:2019qqp}%
  \BibitemOpen
  \bibfield  {author} {\bibinfo {author} {\bibfnamefont {Y.}~\bibnamefont {Kumaran}}\ and\ \bibinfo {author} {\bibfnamefont {A.}~\bibnamefont {\"Ovg\"un}},\ }\bibfield  {title} {\bibinfo {title} {{Weak Deflection Angle of Extended Uncertainty Principle Black Holes}},\ }\href {https://doi.org/10.1088/1674-1137/44/2/025101} {\bibfield  {journal} {\bibinfo  {journal} {Chin. Phys. C}\ }\textbf {\bibinfo {volume} {44}},\ \bibinfo {pages} {025101} (\bibinfo {year} {2020})},\ \Eprint {https://arxiv.org/abs/1905.11710} {arXiv:1905.11710 [gr-qc]} \BibitemShut {NoStop}%
\bibitem [{\citenamefont {Cheng}\ and\ \citenamefont {Zhong}(2021)}]{Cheng:2019zgc}%
  \BibitemOpen
  \bibfield  {author} {\bibinfo {author} {\bibfnamefont {H.}~\bibnamefont {Cheng}}\ and\ \bibinfo {author} {\bibfnamefont {Y.}~\bibnamefont {Zhong}},\ }\bibfield  {title} {\bibinfo {title} {{Instability of a black hole with f (R) global monopole under extended uncertainty principle}},\ }\href {https://doi.org/10.1088/1674-1137/ac1668} {\bibfield  {journal} {\bibinfo  {journal} {Chin. Phys. C}\ }\textbf {\bibinfo {volume} {45}},\ \bibinfo {pages} {105102} (\bibinfo {year} {2021})},\ \Eprint {https://arxiv.org/abs/1908.08201} {arXiv:1908.08201 [hep-th]} \BibitemShut {NoStop}%
\bibitem [{\citenamefont {Hassanabadi}\ \emph {et~al.}(2021)\citenamefont {Hassanabadi}, \citenamefont {Chung}, \citenamefont {L\"utf\"uo\u{g}lu},\ and\ \citenamefont {Maghsoodi}}]{Hassanabadi:2021kyv}%
  \BibitemOpen
  \bibfield  {author} {\bibinfo {author} {\bibfnamefont {H.}~\bibnamefont {Hassanabadi}}, \bibinfo {author} {\bibfnamefont {W.~S.}\ \bibnamefont {Chung}}, \bibinfo {author} {\bibfnamefont {B.~C.}\ \bibnamefont {L\"utf\"uo\u{g}lu}},\ and\ \bibinfo {author} {\bibfnamefont {E.}~\bibnamefont {Maghsoodi}},\ }\bibfield  {title} {\bibinfo {title} {{Effects of a new extended uncertainty principle on Schwarzschild and Reissner\textendash{}Nordstr\"om black holes thermodynamics}},\ }\href {https://doi.org/10.1142/S0217751X21500366} {\bibfield  {journal} {\bibinfo  {journal} {Int. J. Mod. Phys. A}\ }\textbf {\bibinfo {volume} {36}},\ \bibinfo {pages} {2150036} (\bibinfo {year} {2021})}\BibitemShut {NoStop}%
\bibitem [{\citenamefont {Hamil}\ and\ \citenamefont {L\"utf\"uo\u{g}lu}(2021)}]{Hamil:2021ilv}%
  \BibitemOpen
  \bibfield  {author} {\bibinfo {author} {\bibfnamefont {B.}~\bibnamefont {Hamil}}\ and\ \bibinfo {author} {\bibfnamefont {B.~C.}\ \bibnamefont {L\"utf\"uo\u{g}lu}},\ }\bibfield  {title} {\bibinfo {title} {{The effect of higher-order extended uncertainty principle on the black hole thermodynamics}},\ }\href {https://doi.org/10.1209/0295-5075/134/50007} {\bibfield  {journal} {\bibinfo  {journal} {EPL}\ }\textbf {\bibinfo {volume} {134}},\ \bibinfo {pages} {50007} (\bibinfo {year} {2021})}\BibitemShut {NoStop}%
\bibitem [{\citenamefont {\"Okc\"u}\ and\ \citenamefont {Aydiner}(2022)}]{Okcu:2022sio}%
  \BibitemOpen
  \bibfield  {author} {\bibinfo {author} {\bibfnamefont {O.}~\bibnamefont {\"Okc\"u}}\ and\ \bibinfo {author} {\bibfnamefont {E.}~\bibnamefont {Aydiner}},\ }\bibfield  {title} {\bibinfo {title} {{Investigating bounds on the extended uncertainty principle metric through astrophysical tests}},\ }\href {https://doi.org/10.1209/0295-5075/ac6976} {\bibfield  {journal} {\bibinfo  {journal} {EPL}\ }\textbf {\bibinfo {volume} {138}},\ \bibinfo {pages} {39002} (\bibinfo {year} {2022})}\BibitemShut {NoStop}%
\bibitem [{\citenamefont {Hamil}\ \emph {et~al.}(2022)\citenamefont {Hamil}, \citenamefont {L\"utf\"uo\u{g}lu},\ and\ \citenamefont {Dahbi}}]{Hamil:2022bpd}%
  \BibitemOpen
  \bibfield  {author} {\bibinfo {author} {\bibfnamefont {B.}~\bibnamefont {Hamil}}, \bibinfo {author} {\bibfnamefont {B.~C.}\ \bibnamefont {L\"utf\"uo\u{g}lu}},\ and\ \bibinfo {author} {\bibfnamefont {L.}~\bibnamefont {Dahbi}},\ }\bibfield  {title} {\bibinfo {title} {{EUP-corrected thermodynamics of BTZ black hole}},\ }\href {https://doi.org/10.1142/S0217751X22501305} {\bibfield  {journal} {\bibinfo  {journal} {Int. J. Mod. Phys. A}\ }\textbf {\bibinfo {volume} {37}},\ \bibinfo {pages} {2250130} (\bibinfo {year} {2022})},\ \Eprint {https://arxiv.org/abs/2203.09394} {arXiv:2203.09394 [gr-qc]} \BibitemShut {NoStop}%
\bibitem [{\citenamefont {Chen}\ \emph {et~al.}(2022)\citenamefont {Chen}, \citenamefont {Hassanabadi}, \citenamefont {L\"utf\"uo\u{g}lu},\ and\ \citenamefont {Long}}]{Chen:2022ngd}%
  \BibitemOpen
  \bibfield  {author} {\bibinfo {author} {\bibfnamefont {H.}~\bibnamefont {Chen}}, \bibinfo {author} {\bibfnamefont {H.}~\bibnamefont {Hassanabadi}}, \bibinfo {author} {\bibfnamefont {B.~C.}\ \bibnamefont {L\"utf\"uo\u{g}lu}},\ and\ \bibinfo {author} {\bibfnamefont {Z.-W.}\ \bibnamefont {Long}},\ }\bibfield  {title} {\bibinfo {title} {{Quantum corrections to the quasinormal modes of the Schwarzschild black hole}},\ }\href {https://doi.org/10.1007/s10714-022-03037-9} {\bibfield  {journal} {\bibinfo  {journal} {Gen. Rel. Grav.}\ }\textbf {\bibinfo {volume} {54}},\ \bibinfo {pages} {143} (\bibinfo {year} {2022})},\ \Eprint {https://arxiv.org/abs/2203.03464} {arXiv:2203.03464 [gr-qc]} \BibitemShut {NoStop}%
\bibitem [{\citenamefont {Ong}(2020)}]{Ong:2020tvo}%
  \BibitemOpen
  \bibfield  {author} {\bibinfo {author} {\bibfnamefont {Y.~C.}\ \bibnamefont {Ong}},\ }\bibfield  {title} {\bibinfo {title} {{Schwinger pair production and the extended uncertainty principle: can heuristic derivations be trusted?}},\ }\href {https://doi.org/10.1140/epjc/s10052-020-8363-2} {\bibfield  {journal} {\bibinfo  {journal} {Eur. Phys. J. C}\ }\textbf {\bibinfo {volume} {80}},\ \bibinfo {pages} {777} (\bibinfo {year} {2020})},\ \Eprint {https://arxiv.org/abs/2005.12075} {arXiv:2005.12075 [gr-qc]} \BibitemShut {NoStop}%
\bibitem [{\citenamefont {Nozari}\ and\ \citenamefont {Fazlpour}(2007)}]{Nozari:2006wn}%
  \BibitemOpen
  \bibfield  {author} {\bibinfo {author} {\bibfnamefont {K.}~\bibnamefont {Nozari}}\ and\ \bibinfo {author} {\bibfnamefont {B.}~\bibnamefont {Fazlpour}},\ }\bibfield  {title} {\bibinfo {title} {{Can Quantum Gravitational Effects Manifest themselves at Large Distances?}},\ }\href {https://doi.org/10.1016/j.chaos.2006.05.092} {\bibfield  {journal} {\bibinfo  {journal} {Chaos Solitons Fractals}\ }\textbf {\bibinfo {volume} {32}},\ \bibinfo {pages} {1} (\bibinfo {year} {2007})},\ \Eprint {https://arxiv.org/abs/hep-th/0607011} {arXiv:hep-th/0607011} \BibitemShut {NoStop}%
\bibitem [{\citenamefont {Roushan}\ \emph {et~al.}(2024)\citenamefont {Roushan}, \citenamefont {Rashidi},\ and\ \citenamefont {Nozari}}]{Roushan:2024fog}%
  \BibitemOpen
  \bibfield  {author} {\bibinfo {author} {\bibfnamefont {M.}~\bibnamefont {Roushan}}, \bibinfo {author} {\bibfnamefont {N.}~\bibnamefont {Rashidi}},\ and\ \bibinfo {author} {\bibfnamefont {K.}~\bibnamefont {Nozari}},\ }\bibfield  {title} {\bibinfo {title} {{Traces of Quantum Gravity Effects at Late-time Cosmological Dynamics via Distance Measures}},\ }\href {https://doi.org/10.3847/1538-4357/ad74f5} {\bibfield  {journal} {\bibinfo  {journal} {Astrophys. J.}\ }\textbf {\bibinfo {volume} {974}},\ \bibinfo {pages} {263} (\bibinfo {year} {2024})},\ \Eprint {https://arxiv.org/abs/2408.15604} {arXiv:2408.15604 [gr-qc]} \BibitemShut {NoStop}%
\bibitem [{\citenamefont {Lobos}\ and\ \citenamefont {Pantig}(2022)}]{Lobos:2022jsz}%
  \BibitemOpen
  \bibfield  {author} {\bibinfo {author} {\bibfnamefont {N.~J. L.~S.}\ \bibnamefont {Lobos}}\ and\ \bibinfo {author} {\bibfnamefont {R.~C.}\ \bibnamefont {Pantig}},\ }\bibfield  {title} {\bibinfo {title} {{Generalized Extended Uncertainty Principle Black Holes: Shadow and Lensing in the Macro- and Microscopic Realms}},\ }\href {https://doi.org/10.3390/physics4040084} {\bibfield  {journal} {\bibinfo  {journal} {MDPI Physics}\ }\textbf {\bibinfo {volume} {4}},\ \bibinfo {pages} {1318} (\bibinfo {year} {2022})},\ \Eprint {https://arxiv.org/abs/2208.00618} {arXiv:2208.00618 [gr-qc]} \BibitemShut {NoStop}%
\bibitem [{\citenamefont {Pantig}\ \emph {et~al.}(2025)\citenamefont {Pantig}, \citenamefont {Lambiase}, \citenamefont {\"Ovg\"un},\ and\ \citenamefont {Lobos}}]{Pantig:2024asu}%
  \BibitemOpen
  \bibfield  {author} {\bibinfo {author} {\bibfnamefont {R.~C.}\ \bibnamefont {Pantig}}, \bibinfo {author} {\bibfnamefont {G.}~\bibnamefont {Lambiase}}, \bibinfo {author} {\bibfnamefont {A.}~\bibnamefont {\"Ovg\"un}},\ and\ \bibinfo {author} {\bibfnamefont {N.~J. L.~S.}\ \bibnamefont {Lobos}},\ }\bibfield  {title} {\bibinfo {title} {{Spacetime-curvature induced uncertainty principle: Linking the large-structure global effects to the local black hole physics}},\ }\href {https://doi.org/10.1016/j.dark.2025.101817} {\bibfield  {journal} {\bibinfo  {journal} {Phys. Dark Univ.}\ }\textbf {\bibinfo {volume} {47}},\ \bibinfo {pages} {101817} (\bibinfo {year} {2025})},\ \Eprint {https://arxiv.org/abs/2412.00303} {arXiv:2412.00303 [gr-qc]} \BibitemShut {NoStop}%
\bibitem [{\citenamefont {Perlick}\ and\ \citenamefont {Tsupko}(2022)}]{Perlick:2021aok}%
  \BibitemOpen
  \bibfield  {author} {\bibinfo {author} {\bibfnamefont {V.}~\bibnamefont {Perlick}}\ and\ \bibinfo {author} {\bibfnamefont {O.~Y.}\ \bibnamefont {Tsupko}},\ }\bibfield  {title} {\bibinfo {title} {{Calculating black hole shadows: Review of analytical studies}},\ }\href {https://doi.org/10.1016/j.physrep.2021.10.004} {\bibfield  {journal} {\bibinfo  {journal} {Phys. Rept.}\ }\textbf {\bibinfo {volume} {947}},\ \bibinfo {pages} {1} (\bibinfo {year} {2022})},\ \Eprint {https://arxiv.org/abs/2105.07101} {arXiv:2105.07101 [gr-qc]} \BibitemShut {NoStop}%
\bibitem [{\citenamefont {Akiyama}\ \emph {et~al.}(2019{\natexlab{a}})\citenamefont {Akiyama} \emph {et~al.}}]{EventHorizonTelescope:2019dse}%
  \BibitemOpen
  \bibfield  {author} {\bibinfo {author} {\bibfnamefont {K.}~\bibnamefont {Akiyama}} \emph {et~al.} (\bibinfo {collaboration} {Event Horizon Telescope}),\ }\bibfield  {title} {\bibinfo {title} {{First M87 Event Horizon Telescope Results. I. The Shadow of the Supermassive Black Hole}},\ }\href {https://doi.org/10.3847/2041-8213/ab0ec7} {\bibfield  {journal} {\bibinfo  {journal} {Astrophys. J. Lett.}\ }\textbf {\bibinfo {volume} {875}},\ \bibinfo {pages} {L1} (\bibinfo {year} {2019}{\natexlab{a}})},\ \Eprint {https://arxiv.org/abs/1906.11238} {arXiv:1906.11238 [astro-ph.GA]} \BibitemShut {NoStop}%
\bibitem [{\citenamefont {Akiyama}\ \emph {et~al.}(2019{\natexlab{b}})\citenamefont {Akiyama} \emph {et~al.}}]{EventHorizonTelescope:2019ths}%
  \BibitemOpen
  \bibfield  {author} {\bibinfo {author} {\bibfnamefont {K.}~\bibnamefont {Akiyama}} \emph {et~al.} (\bibinfo {collaboration} {Event Horizon Telescope}),\ }\bibfield  {title} {\bibinfo {title} {{First M87 Event Horizon Telescope Results. IV. Imaging the Central Supermassive Black Hole}},\ }\href {https://doi.org/10.3847/2041-8213/ab0e85} {\bibfield  {journal} {\bibinfo  {journal} {Astrophys. J. Lett.}\ }\textbf {\bibinfo {volume} {875}},\ \bibinfo {pages} {L4} (\bibinfo {year} {2019}{\natexlab{b}})},\ \Eprint {https://arxiv.org/abs/1906.11241} {arXiv:1906.11241 [astro-ph.GA]} \BibitemShut {NoStop}%
\bibitem [{\citenamefont {Akiyama}\ \emph {et~al.}(2022{\natexlab{a}})\citenamefont {Akiyama} \emph {et~al.}}]{EventHorizonTelescope:2022xqj}%
  \BibitemOpen
  \bibfield  {author} {\bibinfo {author} {\bibfnamefont {K.}~\bibnamefont {Akiyama}} \emph {et~al.} (\bibinfo {collaboration} {Event Horizon Telescope}),\ }\bibfield  {title} {\bibinfo {title} {{First Sagittarius A* Event Horizon Telescope Results. VI. Testing the Black Hole Metric}},\ }\href {https://doi.org/10.3847/2041-8213/ac6756} {\bibfield  {journal} {\bibinfo  {journal} {Astrophys. J. Lett.}\ }\textbf {\bibinfo {volume} {930}},\ \bibinfo {pages} {L17} (\bibinfo {year} {2022}{\natexlab{a}})},\ \Eprint {https://arxiv.org/abs/2311.09484} {arXiv:2311.09484 [astro-ph.HE]} \BibitemShut {NoStop}%
\bibitem [{\citenamefont {Akiyama}\ \emph {et~al.}(2022{\natexlab{b}})\citenamefont {Akiyama} \emph {et~al.}}]{EventHorizonTelescope:2022wkp}%
  \BibitemOpen
  \bibfield  {author} {\bibinfo {author} {\bibfnamefont {K.}~\bibnamefont {Akiyama}} \emph {et~al.} (\bibinfo {collaboration} {Event Horizon Telescope}),\ }\bibfield  {title} {\bibinfo {title} {{First Sagittarius A* Event Horizon Telescope Results. I. The Shadow of the Supermassive Black Hole in the Center of the Milky Way}},\ }\href {https://doi.org/10.3847/2041-8213/ac6674} {\bibfield  {journal} {\bibinfo  {journal} {Astrophys. J. Lett.}\ }\textbf {\bibinfo {volume} {930}},\ \bibinfo {pages} {L12} (\bibinfo {year} {2022}{\natexlab{b}})},\ \Eprint {https://arxiv.org/abs/2311.08680} {arXiv:2311.08680 [astro-ph.HE]} \BibitemShut {NoStop}%
\bibitem [{\citenamefont {Akiyama}\ \emph {et~al.}(2022{\natexlab{c}})\citenamefont {Akiyama} \emph {et~al.}}]{EventHorizonTelescope:2022wok}%
  \BibitemOpen
  \bibfield  {author} {\bibinfo {author} {\bibfnamefont {K.}~\bibnamefont {Akiyama}} \emph {et~al.} (\bibinfo {collaboration} {Event Horizon Telescope}),\ }\bibfield  {title} {\bibinfo {title} {{First Sagittarius A* Event Horizon Telescope Results. III. Imaging of the Galactic Center Supermassive Black Hole}},\ }\href {https://doi.org/10.3847/2041-8213/ac6429} {\bibfield  {journal} {\bibinfo  {journal} {Astrophys. J. Lett.}\ }\textbf {\bibinfo {volume} {930}},\ \bibinfo {pages} {L14} (\bibinfo {year} {2022}{\natexlab{c}})},\ \Eprint {https://arxiv.org/abs/2311.09479} {arXiv:2311.09479 [astro-ph.HE]} \BibitemShut {NoStop}%
\bibitem [{\citenamefont {Li}\ \emph {et~al.}(2020)\citenamefont {Li}, \citenamefont {Zhang},\ and\ \citenamefont {\"Ovg\"un}}]{Li:2020wvn}%
  \BibitemOpen
  \bibfield  {author} {\bibinfo {author} {\bibfnamefont {Z.}~\bibnamefont {Li}}, \bibinfo {author} {\bibfnamefont {G.}~\bibnamefont {Zhang}},\ and\ \bibinfo {author} {\bibfnamefont {A.}~\bibnamefont {\"Ovg\"un}},\ }\bibfield  {title} {\bibinfo {title} {{Circular Orbit of a Particle and Weak Gravitational Lensing}},\ }\href {https://doi.org/10.1103/PhysRevD.101.124058} {\bibfield  {journal} {\bibinfo  {journal} {Phys. Rev. D}\ }\textbf {\bibinfo {volume} {101}},\ \bibinfo {pages} {124058} (\bibinfo {year} {2020})},\ \Eprint {https://arxiv.org/abs/2006.13047} {arXiv:2006.13047 [gr-qc]} \BibitemShut {NoStop}%
\bibitem [{\citenamefont {Vagnozzi}\ \emph {et~al.}(2023)\citenamefont {Vagnozzi} \emph {et~al.}}]{Vagnozzi:2022moj}%
  \BibitemOpen
  \bibfield  {author} {\bibinfo {author} {\bibfnamefont {S.}~\bibnamefont {Vagnozzi}} \emph {et~al.},\ }\bibfield  {title} {\bibinfo {title} {{Horizon-scale tests of gravity theories and fundamental physics from the Event Horizon Telescope image of Sagittarius A}},\ }\href {https://doi.org/10.1088/1361-6382/acd97b} {\bibfield  {journal} {\bibinfo  {journal} {Class. Quant. Grav.}\ }\textbf {\bibinfo {volume} {40}},\ \bibinfo {pages} {165007} (\bibinfo {year} {2023})},\ \Eprint {https://arxiv.org/abs/2205.07787} {arXiv:2205.07787 [gr-qc]} \BibitemShut {NoStop}%
\bibitem [{\citenamefont {Kocherlakota}\ \emph {et~al.}(2021)\citenamefont {Kocherlakota} \emph {et~al.}}]{EventHorizonTelescope:2021dqv}%
  \BibitemOpen
  \bibfield  {author} {\bibinfo {author} {\bibfnamefont {P.}~\bibnamefont {Kocherlakota}} \emph {et~al.} (\bibinfo {collaboration} {Event Horizon Telescope}),\ }\bibfield  {title} {\bibinfo {title} {{Constraints on black-hole charges with the 2017 EHT observations of M87*}},\ }\href {https://doi.org/10.1103/PhysRevD.103.104047} {\bibfield  {journal} {\bibinfo  {journal} {Phys. Rev. D}\ }\textbf {\bibinfo {volume} {103}},\ \bibinfo {pages} {104047} (\bibinfo {year} {2021})},\ \Eprint {https://arxiv.org/abs/2105.09343} {arXiv:2105.09343 [gr-qc]} \BibitemShut {NoStop}%
\bibitem [{\citenamefont {Chen}\ \emph {et~al.}(2024)\citenamefont {Chen}, \citenamefont {Li}, \citenamefont {Zhu},\ and\ \citenamefont {Wu}}]{Chen:2023bao}%
  \BibitemOpen
  \bibfield  {author} {\bibinfo {author} {\bibfnamefont {R.-T.}\ \bibnamefont {Chen}}, \bibinfo {author} {\bibfnamefont {S.}~\bibnamefont {Li}}, \bibinfo {author} {\bibfnamefont {L.-G.}\ \bibnamefont {Zhu}},\ and\ \bibinfo {author} {\bibfnamefont {J.-P.}\ \bibnamefont {Wu}},\ }\bibfield  {title} {\bibinfo {title} {{Constraints from Solar System tests on a covariant loop quantum black hole}},\ }\href {https://doi.org/10.1103/PhysRevD.109.024010} {\bibfield  {journal} {\bibinfo  {journal} {Phys. Rev. D}\ }\textbf {\bibinfo {volume} {109}},\ \bibinfo {pages} {024010} (\bibinfo {year} {2024})},\ \Eprint {https://arxiv.org/abs/2311.12270} {arXiv:2311.12270 [gr-qc]} \BibitemShut {NoStop}%
\bibitem [{\citenamefont {Fomalont}\ \emph {et~al.}(2009)\citenamefont {Fomalont}, \citenamefont {Kopeikin}, \citenamefont {Lanyi},\ and\ \citenamefont {Benson}}]{Fomalont_2009}%
  \BibitemOpen
  \bibfield  {author} {\bibinfo {author} {\bibfnamefont {E.}~\bibnamefont {Fomalont}}, \bibinfo {author} {\bibfnamefont {S.}~\bibnamefont {Kopeikin}}, \bibinfo {author} {\bibfnamefont {G.}~\bibnamefont {Lanyi}},\ and\ \bibinfo {author} {\bibfnamefont {J.}~\bibnamefont {Benson}},\ }\bibfield  {title} {\bibinfo {title} {Progress in measurements of the gravitational bending of radio waves using the vlba},\ }\href {https://doi.org/10.1088/0004-637X/699/2/1395} {\bibfield  {journal} {\bibinfo  {journal} {The Astrophysical Journal}\ }\textbf {\bibinfo {volume} {699}},\ \bibinfo {pages} {1395} (\bibinfo {year} {2009})}\BibitemShut {NoStop}%
\end{thebibliography}%

\end{document}